\documentclass[preprint]{imsart}

\RequirePackage[OT1]{fontenc}
\RequirePackage{amsthm,amsmath,amssymb}
\RequirePackage[numbers]{natbib}
\RequirePackage[colorlinks,citecolor=blue,urlcolor=blue]{hyperref}
\RequirePackage{graphicx}
\usepackage{hyperref}
\usepackage{algorithm}
\usepackage{algpseudocode}
\usepackage{mathtools}

\startlocaldefs
\numberwithin{equation}{section}
\theoremstyle{plain}

\newcommand{\mc}{\mathcal}
\newcommand{\mb}{\mathbb}

\newcommand{\oss}{\mathcal{O}_{\tau ,s}}

\endlocaldefs

\begin{document}
\begin{frontmatter}
\title{The Bayesian SLOPE}
\runtitle{The Bayesian SLOPE}

\begin{aug}
\author{\fnms{Amir} \snm{Sepehri}\ead[label=e1]{asepehri@stanford.edu}}

\address{Department of Statistics\\
390 Serra Mall\\
Stanford University\\
Stanford, CA 94305-4065\\
\printead{e1}}

\runauthor{A. Sepehri}
\affiliation{Stanford University}

\end{aug}

\begin{abstract} 
The SLOPE \cite{bogdan2015slope,figueiredo2014sparse} estimates regression coefficients by minimizing a regularized residual sum of squares using a sorted-$\ell_1$-norm penalty. The SLOPE combines testing and estimation in regression problems. It exhibits suitable variable selection and prediction properties, as well as minimax optimality.
This paper introduces the Bayesian SLOPE procedure for linear regression. The classical SLOPE estimate is the posterior mode in the normal regression problem with an appropriate prior on the coefficients. The Bayesian SLOPE considers the full Bayesian model and has the advantage of offering credible sets and standard error estimates for the parameters. Moreover, the hierarchical Bayesian framework allows for full Bayesian and empirical Bayes treatment of the penalty coefficients; whereas it is not clear how to choose these coefficients when using the SLOPE on a general design matrix.  A direct characterization of the posterior is provided which suggests a Gibbs sampler that does not involve latent variables. An efficient hybrid Gibbs sampler for the Bayesian SLOPE is introduced. Point estimation using the posterior mean is highlighted, which automatically facilitates the Bayesian prediction of future observations. These are demonstrated on real and synthetic data. Implementation of the Bayesian SLOPE in R is provided as supplementary material \ref{SM}.
\end{abstract}



\end{frontmatter}
\section{Introduction}\label{lntro}
Consider estimating $\beta$ in the linear regression model 
\begin{align*}
y = X\beta +\epsilon,
\end{align*} 
where $y$ is an $n\times 1$ response vector, $X$ an $n \times p$ (standardized) design matrix, $\beta$ the $p\times 1$ vector of regression coefficients, and $\epsilon$ an $n\times 1$ vector of independent normal errors with mean $0$ and variance $\sigma^2$. The SLOPE estimate is the solution to the following regularized least squares regression problem:
\begin{align}\label{Def:SLOPE}
\min_{\beta \in \mb{R}^p} \frac{1}{2} \|y-X\beta\|^2_{\ell_2}+\sigma \sum_{i=1}^p \lambda_i |\beta|_{(i)},
\end{align}
where $|\beta|_{(1)} \ge \ldots\ge |\beta|_{(p)} $ are the absolute values of the entries of $\beta$ in  decreasing order and $\lambda_1\ge  \ldots\ge \lambda_p\ge 0$ are tuning parameters (the vector of penalty coefficients). The SLOPE procedure provides a bridge between the lasso estimation procedure \citep{tibshirani1996regression} and \textit{false discovery rate} (FDR) controling multiple testing procedures such as the Benjamini-Hochberg procedure (BHq) \citep{benjamini1995controlling}. It uses the sorted $\ell_1$ penalty which generalizes the $\ell_1$ regularization used in lasso, by penalizing larger coefficients more stringently. Penalizing larger coefficients more stringently is similar to BHq, which compares more significant $p$-values with more stringent thresholds. In fact, the SLOPE has been shown to control the FDR for orthogonal design matrices \cite{bogdan2015slope}, and produces sparse vector of regression coefficients. We refer the reader to \cite{bogdan2015slope,figueiredo2014sparse,su2015slope} for further details about the SLOPE and its properties.

Representation \ref{Def:SLOPE} suggests that the SLOPE estimate can be derived as the maximum a posteriori of $\beta$ in a Bayesian regression model, defined as follows. Define the SLOPE prior $\pi(\beta\mid \sigma^2, \lambda)$ as
\begin{align}\label{def:SLOPEprior}
\pi(\beta\mid \sigma^2, \lambda) = C(\lambda,\sigma^2) e^{\frac{-1}{\sigma} \sum_{i=1}^p \lambda_i |\beta|_{(i)}},
\end{align}
where $C(\lambda,\sigma^2)$ is the appropriate normalizing constant. As shown in appendix \ref{NormalizingCons}, $C(\lambda,\sigma^2)$ is
\begin{align*}
C(\lambda,\sigma^2) = \frac{\lambda_1(\lambda_1+\lambda_2)\ldots (\lambda_1+\lambda_2+\ldots+\lambda_p)}{2^p \sigma^p p! }.
\end{align*}
With this notation, the Bayesian SLOPE regression model is defined as
\begin{align}\label{BayesModel}
\begin{split}
y\mid \beta,\sigma^2 &\sim \mc{N}(X\beta,\sigma^2 I),\\
\pi(\beta\mid \sigma^2, \lambda) &= C(\lambda,\sigma^2) e^{\frac{-1}{\sigma} \sum_{i=1}^p \lambda_i |\beta|_{(i)}},
\end{split}
\end{align}
where independent priors $\pi(\sigma^2)$ and $\pi(\lambda)$ can be assumed on $\sigma^2$ and $\lambda$, respectively. The choice of prior on hyper-parameters and the posterior distribution are discussed in Section \ref{sec:Posterior}. The SLOPE estimate is then the maximum a posteriori for $\beta$ in this model, conditional on $\sigma^2$ and $\lambda$. 

\textit{Remark.} Alternatively, one can define of the SLOPE estimate as the solution to the following regularized regression problem:
\begin{align*}
\min_{\beta \in \mb{R}^p} \frac{1}{2} \|y-X\beta\|^2_{\ell_2}+ \sum_{i=1}^p \lambda_i |\beta|_{(i)},
\end{align*}
where scaling of the penalty on $\beta$ does not depend $\sigma$. However, we choose not to pursue this path because of the difficulties posed by the possibility of a non-unimodal posterior for $\beta$. A multi-modal posterior causes conceptual and computational difficulties. It is challenging to summarize a multi-modal posterior with a single point estimate, as any reasonable summary needs to provide information about different modes along with a measure of the corresponding probability mass around each mode. Furthermore, a multi-modal target distribution can slow the Markov chain Monte Carlo methods to a prohibitive extent.  For a discussion of the issues related to use of this prior in the Bayesian lasso problem, as well as an example of a multi-modal posterior, see Section 4 of \cite{park2008bayesian}. It is seen in Appendix \ref{AppA} that using the formulation (\ref{Def:SLOPE}) has the advantage of producing a unimodal joint posterior distribution for $(\beta,\sigma^2)$.

There is a sizable literature on Bayesian interpretation of regularized regression methods, including the Bayesian lasso \citep{hans2009bayesian,hans2010model,park2008bayesian}, the Bayesian Elastic Net \citep{bornn2010grouping,hans2011elastic,li2010bayesian}, the Bayesian group lasso \citep{xu2015bayesian}, the Bayesian Bridge \cite{polson2014bayesian}, and the Bayesian regularized quantile regression \citep{li2010bayesian2}. There is also a vast literature on the closely related topic of Bayesian variable selection in linear regression. Examples include, but not limited to, the Spike and Slab variable selection and its variants \citep{hernandez2013generalized,ishwaran2005spike,ishwaran2011consistency,rockova2014spike,rockova2015bayesian,yen2011majorization}, variational methods such as Expectation-Maximization variable selection \citep{carbonetto2012scalable,rovckova2014emvs,yen2011majorization}, the Horseshoe estimator \citep{carvalho2010horseshoe,van2014horseshoe}, and many other methods \citep{bhattacharya2015dirichlet,castillo2015bayesian,martin2014asymptotically,polson2010shrink,raykar2010nonparametric,scott2010bayes}. Consistency and optimality of some of these methods have been studied in \citep{bhattacharya2016sub,castillo2012needles,ishwaran2011consistency,martin2014asymptotically,moreno2015posterior,rovckova2016bayesian,van2014horseshoe}.
Particular attention has been paid to the optimality properties in the minimax sense. Results along these lines include proof of minimax optimality for posterior mode or posterior mean. Minimax optimality for the posterior mode of the Bayesian SLOPE , i.e.\ the SLOPE estimate, has been already shown in \cite{su2015slope} for a random design matrix, and in \cite{bellec2016slope} for a general design matrix under a Restricted Eigenvalue type condition. 

Most of the regularized regression methods use separable penalties, that are sums of individual penalties for each coefficient, which correspond to independent priors on the coefficient vector. On the other hand, many of the Bayesian variable selection methods mentioned above use hidden model structures which explicitly incorporate variable selection into the Bayesian analysis and, as a byproduct, put non-separable priors on the coefficient vector. Non-separable priors capture the global structure of the coefficient vector better than separable priors; see \cite{rovckova2016bayesian} for a further discussion. However, hidden model structure may slow down the posterior sampling significantly, as the they need to sample from a distribution in higher dimensions to account for the latent variables encoding the hidden structure. Depending on the problem in hand, it may be unsatisfying to assume an underlying model in which some coefficients can be exactly zero. Another approach is to carry out full Bayesian analysis using a prior, e.g.\ the SLOPE prior, on the coefficients. The Bayesian SLOPE benefits from a non-separable prior, which captures the global features of $\beta$, as well as a log-concave posterior, which allows for much faster sampling of the posterior.

This paper formulates the Bayesian SLOPE, offering a full Bayesian analogue of the SLOPE procedure.  A direct characterization of the posterior distribution $\pi(\beta\mid y, \sigma^2, \lambda)$ is introduced in Section \ref{sec:Posterior}, followed by a discussion of estimation and prediction under the SLOPE prior from a Bayesian model-based perspective. Particularly, prediction via the posterior predictive distribution is discussed and compared with the SLOPE prediction.
The direct characterization of the posterior is used to design a Gibbs sampler without using latent variables. A Hamiltonian Monte Carlo samplers is introduced which can be faster than the Gibbs sampler. This is discussed in Section \ref{MCMC}. Bayesian and empirical Bayes treatment of the vector of tuning parameters, $\lambda$, is discussed in Section \ref{PenaltyVector}. Application of these methods on simulated and real world examples are presented in Section \ref{Ex}.

\section{The SLOPE posterior distribution} \label{sec:Posterior}
\subsection{Piecewise normal characterization of the posterior}
The posterior distribution of the vector of coefficients equals
\begin{align}\label{eqn:PosteriorBeta}
\pi(\beta\mid y, \sigma^2, \lambda)  \propto e^{- \frac{1}{2\sigma^2} \|y- X\beta\|^2 - \frac{1}{\sigma} \sum_{i=1}^p \lambda_i |\beta|_{(i)}},
\end{align} 
which is proportional to the density of a multivariate normal distribution for any fixed order of $\{|\beta_i| ; i = 1,\ldots,p \}$ and signs of the coefficients $\{\beta_i; i = 1,\ldots,p \}$. To make the statement precise, for a permutation $\tau \in \mc{S}_p$ and a sign vector $s\in \{\pm 1\}^p$, define 
\begin{align*}
\mc{O}_{\tau ,s} = \{\beta \in \mb{R}^p\mid sign(\beta_i) = s_i \; , |\beta_{\tau(1)}|\ge \ldots \ge |\beta_{\tau(p)}|\ge 0\},
\end{align*}
 where $\mc{S}_p$ is the group of all permutations of the set $\{1,\ldots,p\}$
The posterior can be written as
\begin{align*}
\pi(\beta\mid y, \sigma^2, \lambda)  &\propto \sum_{\tau\in \mc{S}_p, s\in \{\pm 1\}^p } e^{- \frac{1}{2\sigma^2} \|y- X\beta\|^2 - \frac{1}{\sigma} \sum_{i=1}^p \lambda_i s_{\tau(i)} \beta_{\tau(i)}} \mb{I}_{\beta \in \mc{O}_{\tau,s}},
\end{align*}
which is a weighted sum of multivariate normal densities each restricted to one of the sets $\oss$ for $\tau \in \mc{S}_p$ and $s \in \{\pm 1\}^p$.
Denote by $\mc{N}^{\tau,s}(x \mid \mu, \Sigma)$ the multivariate normal density with mean vector $\mu$ and covariance matrix $\Sigma$, truncated to $\oss$. The posterior can be written as
\begin{align}\label{eqn:DirectPosterior}
\pi(\beta\mid y, \sigma^2, \lambda)  &= \sum_{\tau \in \mc{S}_p, s\in \{\pm 1\}^p } w_{\tau ,s}\, \mc{N}^{\tau ,s}(\beta \mid \mu_{\tau ,s}, \Sigma),
\end{align}
with the common covariance structure $\Sigma = \sigma^2 (X^TX)^{-1}$ and the orthant-dependent means and weights
\begin{align*}
\mu_{\tau,s} = \hat{\beta}_{OLS} - \frac{1}{\sigma}\Sigma D_{\tau ,s} \lambda,\;\;
w_{\tau,s}     = \frac{e^{\frac{1}{2}\mu_{\tau ,s}^T \Sigma^{-1}\mu_{\tau,s}}}{\sum_{\pi \in \mc{S}_p,  r\in \{\pm 1\}^p } e^{\frac{1}{2}\mu_{\pi,r}^T \Sigma^{-1}\mu_{\pi,r}} m_{\pi,r}},
\end{align*}
where $\hat{\beta}_{OLS} = (X^TX)^{-1}X^Ty$ is the ordinary regression coefficient vector, $D_{\tau ,s}$ is the signed permutation matrix corresponding to the permutation $\tau$ and signs vector $s$, and $m_{\tau ,s} = \int \mc{N}^{\tau ,s}(\beta \mid \mu_{\tau,s}, \Sigma) d \beta$.

The model can be extended with specifying priors on variance of the noise. A typical choice for the prior on $\sigma^2$ is the inverse gamma prior
\begin{align} \label{Def:PriorSigma}
\pi(\sigma^2) = \frac{\gamma^a}{\Gamma(a)} (\sigma^2)^{-a-1} e^{-\gamma/\sigma^2}.
\end{align} 
The model (\ref{BayesModel}), along with (\ref{Def:PriorSigma}), define a full Bayesian regression model with hyper-parameters $a, \gamma, \text{and }\lambda$.
The full posterior can be sampled using Markov chain Monte Carlo methods discussed in Section \ref{MCMC}.

\textit{Remark.} Instead of the prior (\ref{Def:PriorSigma}) on $\sigma^2$, one can use the non-informative improper prior $\pi(\sigma^2) \propto 1/\sigma^2$, which is a special case of (\ref{Def:PriorSigma}) with $a=\gamma =0$. This choice of prior induces a proper posterior and the joint posterior for $(\beta,\sigma^2)$ is again unimodal, which can be sampled similarly to the posterior resulting from (\ref{Def:PriorSigma}).

The posterior distribution of $(\beta,\sigma^2)$ is usually the main object of interest in a Bayesian regression problem. However, one might carry out a Bayesian analysis about the regularization coefficients too, to take into account other types of prior information available. Choosing a reasonable prior on $\lambda$ depends on information the practitioner has. A conjugate prior is proposed in Section \ref{HyperLambda}. Empirical Bayes choice of $\lambda$ is discussed in Section $\ref{EmpVarBay}$.
\subsection{Estimation and prediction based on the posterior}
Two major tasks of interest in linear regression problems are point estimation of the parameters and prediction of the response for future observations. The Bayesian point estimate of $\beta$, under a given loss function $\ell(\hat{\beta}, \beta)$, is the estimator $\hat{\beta}$ minimizing the expected posterior loss, $\int \ell(\hat{\beta},\beta)\pi(\beta\mid \sigma^2,\lambda,y)d \beta$. Common choices are the posterior mean and median, which are the point estimates corresponding to squared-error loss and absolute-error loss functions, respectively. The SLOPE estimate, $\hat{\beta}_{SLOPE}$, corresponds to the posterior mode. Although using the posterior mode as a Bayesian point estimate has become more popular recently, it seems to be an unnatural choice for a Bayesian statistician. Particularly, it can be realized as the $\epsilon \downarrow 0$ limit of Bayes estimates corresponding to loss functions $1- \mb{I}_{\|\hat{\beta}-\beta\|<\epsilon}$. Although choosing the loss function is subjective and up to the statistician, this choice of loss function seems rather unnatural.

Equally important is the task of predicting the response for new observations. Consider a new observation $X_0$ at which one wishes to predict the response. The Bayesian prediction of the future value is made using the posterior predictive distribution,
\begin{align*}
p(y_0\mid \sigma^2,\lambda, y) = \int p(y_0\mid \beta, \sigma^2,\lambda, y) \pi(\beta \mid \sigma^2, \lambda, y)d \beta.
\end{align*}
For a loss function $\ell(\tilde{y},y_0)$, the Bayesian prediction is based on the predictor $\tilde{y}$ minimizing the expected posterior predictive loss,
\begin{align*}
R(\tilde{y},y_0) = \int \ell(\tilde{y},y_0) p(y_0\mid \sigma^2, \lambda, y)d y_0.
\end{align*}
Under the squared-error loss the prediction is done using the mean of the posterior predictive distribution, given by $\tilde{y} = X_0 \mb{E}(\beta \mid \sigma^2, \lambda, y)$. An important advantage of the squared-error loss is the fact that the posterior mean provides both point estimation and prediction. On the other hand, the mode of the posterior predictive distribution, $p(y_0\mid \sigma^2,\lambda, y)$, is not equal to $X_0 \hat{\beta}_{SLOPE}$. An example in which this is the case for the univariate lasso problem is provided in \cite{hans2009bayesian}. The popular prediction rule given by $\tilde{y} = X_0 \hat{\beta}_{SLOPE}$, although useful, does not seem to have a solid Bayesian justification. The posterior mean is a more natural choice for prediction.

\section{Markov chain Monte Carlo sampling from posterior}\label{MCMC}
\subsection{The standard Gibbs sampler}\label{Gibbs}
The Gibbs sampler is the most commonly used sampling method in Bayesian analysis. Most of the Bayesian variable selection methods mentioned in Section \ref{lntro} use Gibbs sampling to sample from the posterior.  A Gibbs sampler for the SLOPE posterior, which updates each parameter on at a time, is described in this Section. The direct characterization of the posterior, (\ref{eqn:DirectPosterior}), is used to compute the conditional posterior for $\beta_j$, which is piecewise normal. For a fixed $j$, let $x_1 \ge \ldots \ge x_{p-1}$ be the sorted values of $\{|\beta_i| \mid i \neq j\}$. For $k = 1,\ldots,p$, let $\mc{N}^{k}(.\mid \mu, \eta^2)$ and $\mc{N}^{-k}(.\mid \mu, \eta^2)$ be the normal density with mean $\mu$ and variance $\eta^2$ truncated to $[x_{k},x_{k-1})$ and $(-x_{k-1},-x_{k}]$, respectively, where $x_0=\infty$ and $x_p =0$. With this notation, the conditional posterior distributions are
\begin{align}
\pi(\beta_j \mid \beta_{-j},\sigma^2,\lambda,y) &= \sum_{s= \pm 1} \sum_{k=1}^p  \phi_{j,s k} \, \mc{N}^{s k}(\beta_j \mid \mu_{j,s k}, \omega_{jj}^{-1}), \label{betaPostCond}\\
\pi(\sigma^2 \mid \beta,\lambda,y)					   &\propto (\sigma^2)^{-a^*-1} e^{-\gamma^*/\sigma^2 - \alpha^*/\sigma}.\label{sigmaPostCond}
\end{align}
The weights and means in (\ref{betaPostCond}) are (for $  s=\pm 1,\,\, k=1,\ldots,p $) 
\begin{align}
\mu_{j,sk} &= \hat{\beta}_{OLS,j}+ \sum_{i \neq j} \frac{\omega_{ij}}{\omega_{jj}}(\hat{\beta}_{OLS,i}-\beta_i) - \frac{s \lambda_k}{\sigma \omega_{jj}},\label{betaMeans}\\
\phi_{j,sk} &= \frac{e^{\mu^2_{j,sk}\,\omega_{jj}/2}}{\sum_{t = \pm 1} \sum_{l=1}^p e^{\mu^2_{j,tl}\,\omega_{jj}/2}\left[\Phi\left(\sqrt{\omega_{jj}}(x_{l-1}-t\mu_{j,tl})\right)-\Phi\left(\sqrt{\omega_{jj}}(x_{l}-t\mu_{j,tl})\right)\right]},\label{betaWeights}
\end{align}
where $\omega_{ij}$ is the $ij$ entry of $\Sigma^{-1}$ . The parameters in (\ref{sigmaPostCond}) are
\begin{align*}
a^* = (n+p)/2+a ,\; \gamma^* = \frac{1}{2} \|y- X\beta\|^2+\gamma, \text{ and }\alpha^* = \sum_{i=1}^p \lambda_i |\beta|_{(i)} .
\end{align*}

The conditional posterior for $\beta_j$ can be sampled using the piecewise normal characterization (\ref{betaPostCond}). Since the mean parameters in (\ref{betaMeans}) change only slightly at each iteration, we only need to update the previous values, which requires linear number of operations in $p$. The weights in (\ref{betaWeights}) can be updated in linear time too, thus, each run through the entire vector $\beta$ requires quadratic number of operations. Thus, the Gibbs sampler is affordable for moderately large $p$. Sampling from the conditional distribution of $\sigma^2$ is discussed in the appendix of \cite{hans2009bayesian}.

The Gibbs sampler can be initialized at $(\beta_{in}, \sigma^2_{in}) = (\hat{\beta}_{SLOPE}, \hat{\sigma}^2)$, where $\hat{\beta}_{SLOPE}$ is the SLOPE estimate and $\hat{\sigma}^2$ is an estimate of the variance from the data. A systematic scan can be used, sampling in the following order: $\beta_j$ for $j=1,2,\ldots,p$ and then $\sigma^2$.  

Although implementing the standard Gibbs sampler is straightforward, in some cases, e.g. when the predictor variables are highly correlated, it can suffer from high autocorrelation. Another limitation, in a large $p$ setting, is the relatively high cost of sampling the conditional distribution for $\beta_j$. Despite the complicated posterior $\pi(\beta \mid \sigma^2,\lambda,y)$, the usual block-updating solution is feasible, thanks to recent developments in Markov chain Monte Carlo simulation. This is presented in Section \ref{HybridGibbs}.

\subsection{An efficient block-updating Gibbs sampler using Hamiltonian Monte Carlo}\label{HybridGibbs}
 The Gibbs sampler from Section \ref{Gibbs} can be improved to a block-updating Gibbs sampler using the Hamiltonian Monte Carlo \citep{duane1987hybrid,neal2011mcmc}, to sample directly from the multivariate conditional distribution $\pi(\beta\mid \sigma^2,\lambda,y)$. To sample from a distribution $p(x) = e^{-U(x)}$ on $\mb{R}^p$, Hamiltonian Monte Carlo expands the parameter space by adding a `momentum' variable $v\in \mb{R}^p$. It samples the momentum from the standard Gaussian distribution and evolves the current state $(x,v)$ by running the Hamiltonian dynamics
 \begin{align*}
 \frac{d x}{dt} = v,\;\; \frac{dv}{dt} = - \dot{U}(x),
 \end{align*}
 with initial condition $(x_0,v_0)$. After a fixed time $T$, the location component $x_T$ is kept and the momentum component $v_T$ is re-sampled. In most applications the Hamilton equations are not exactly solvable; hence a numerical approximation is needed. The most popular numerical method is the \textit{leapfrog} procedure. To account for the approximation error, a Metropolis-Hasting correction is usually used, see \cite{neal2011mcmc} for more details. Hamiltonian Monte Carlo is implemented efficiently in the software system STAN \citep{carpenter2016stan}.
 
It might be possible to improve upon the generic Hamiltonian Monte Carlo implementations by avoiding the rejections from the Metropolis-Hasting filter. \citet{pakman2014exact} provide exact solutions of the Hamilton equations for the case of the truncated (multivariate) normal distribution. This method can be directly used for the SLOPE posterior $\pi(\beta\mid \sigma^2, \lambda,y)$. There is slight subtlety because of the non-smoothness of the posterior for $\beta$, i.e.\ lack of differentiablity at $\beta_i=0$ and $\beta_i = \beta_j$. \citet{chaari2014hamiltonian} have addressed this issue by introducing a Hamiltonian Monte Carlo for non-smooth log-densities, which uses sub-gradients instead of gradients. See \cite{pakman2014exact,chaari2014hamiltonian} for details.

Algorithm \ref{BlockGibbs} describes a block-updating Gibbs sampler based on Hamiltonian Monte Carlo, which can be implemented in the STAN modeling language. A sampler based on Hamiltonian Monte Carlo is implemented in STAN and is available as online supplement, which also provides the R functions required to run Algorithm \ref{BlockGibbs}.
\begin{algorithm}[H]
\caption{The block-updating Gibbs Sampler} \label{BlockGibbs}
\begin{algorithmic}[1]
   \item[0:] Fix $T$.
   \item Initialize the parameters $(\beta_{[0]}, \sigma^2_{[0]}) = (\hat{\beta}_{SLOPE}, \hat{\sigma}^2)$.
   \item Run the Hamiltonian Monte Carlo for time $T$, to sample $\beta_{[k]}$ from $\pi(\beta\mid \sigma^2_{[k-1]},\lambda,y)$.
   \item Sample $\sigma^2_{[k]}$ from $\pi(\sigma^2\mid \beta_{[k]},\lambda,y)$.
   \item Repeat 2 and 3 until convergence.
\end{algorithmic}
\end{algorithm}

\section{Choosing the penalty vector $\lambda$}\label{PenaltyVector}
\subsection{Empirical Bayes estimates for $\lambda$} \label{EmpVarBay}
The model defined by  (\ref{BayesModel}) and (\ref{Def:PriorSigma}) induces a likelihood function for $\lambda$. This likelihood function, computed on the observed data $(X,y)$, can be used to obtain a frequentist estimate of $\lambda$ via Expectation-Maximization (EM) algorithm. In general, for almost all problems, there is no guarantee that the EM algorithm converges to the maximum likelihood estimator, but it increases the likelihood at each step.
The full-data log-likelihood is
\begin{align*}
\ell(y,\beta,\sigma,\lambda) = &\frac{-(\|y-X\beta\|^2+\gamma)}{\sigma^2} -\left(\frac{n+p}{2}+a+1\right)\log (\sigma^2) \\
&- \frac{\sum_{i=1}^p \lambda_i |\beta|_{(i)}}{\sigma} +\sum_{i=1}^p \log\left(\sum_{j=1}^i\lambda_j\right) + \log\left(\mb{I}_{\lambda_1\ge \ldots\ge \lambda_p\ge0} \right).
\end{align*}
The E-step in the EM algorithm computes the expected value of this log-likelihood given $y$, under the distribution with current iterate $\lambda^k$, to get
\begin{align*}
Q(\lambda\mid \lambda^k) = \sum_{i=1}^p \log\left(\sum_{j=1}^i\lambda_j\right) +    &   \log\left(\mb{I}_{\lambda_1\ge \ldots\ge \lambda_p\ge0} \right)- \sum_{i=1}^p \lambda_i \mb{E}_{\lambda^{k-1}}\left[ |\beta|_{(i)}/\sigma\mid y\right] \\ 
&+ \text{terms not involving } \lambda.
\end{align*}
The M-step maximizes $Q(\lambda\mid \lambda^k)$ over $\lambda$ to update the iterate to $\lambda^{k+1} = \text{arg} \max_\lambda Q(\lambda\mid \lambda^k)$. This is a convex optimization problem in $\lambda$ and can be solved efficiently using gradient decent and alternating direction method of multipliers . The EM algorithm is repeated until a desired level of convergence is obtained, i.e.\ $ \|\lambda^{k-1}-\lambda^k\|<\epsilon$. For the Bayesian SLOPE, the EM algorithm is hard to carry out, as there is no analytical expression for $\mb{E}_{\lambda^{k-1}}\left[ |\beta|_{(i)}/\sigma\mid y\right]$. The expectations in the E-step can be computed using Monte Carlo methods; this procedure is called the Monte Carlo EM algorithm \cite{casella2001empirical}. For the Bayesian SLOPE, the steps are described in algorithm \ref{alg:MC-EM}.
\begin{algorithm}[H]
\caption{The Monte Carlo EM algorithm} \label{alg:MC-EM}
\begin{algorithmic}[1]
   \item[0:] Initialize the parameter $\lambda$, e.g\ $\lambda^0=\lambda_{BH}$.
   \item For $k=1,2,\ldots$ repeat
   \item Generate a sample from the posterior distribution of $\beta,\sigma^2$ using the Monte Carlo sampler of Section \ref{MCMC} with $\lambda$ set to $\lambda^{k-1}$.
   \item \textbf{E step} Approximate $ Q(\lambda\mid \lambda^{k-1})$ by substituting $ \mb{E}_{\lambda^{k-1}}\left[ |\beta|_{(i)}/\sigma\mid y\right]$ with the average based on the Monte Carlo sample of step 2, to get $ \widehat{Q}(\lambda\mid \lambda^{k-1})$.
   \item  \textbf{M step} Update the estimate $ \lambda^{k} = \text{arg}\max_{\lambda} \widehat{Q}(\lambda\mid \lambda^{k-1})$.
   \item Break if\ $\|\lambda^{k-1}-\lambda^k\|<\epsilon$.
   \item Output $\lambda^{k}$.
\end{algorithmic}
\end{algorithm}

\subsection{Hyperpriors on $\lambda$}\label{HyperLambda}
This Section considers a Bayesian treatment of the penalty parameter, $\lambda$. It is indeed essential to incorporate any educated suggestion and prior knowledge into the prior distribution of $\lambda$. In the case there is not much known a priori, a generic proposal can be used. For a set of parameters $b_1,\ldots, b_p$ and $c_1,\ldots, c_p$, define
\begin{align}\label{Def:PriorLambda}
\pi(\lambda) \propto e^{- \sum_{i=1}^p b_i \lambda_i} \prod_{i=1}^p (\lambda_1+\ldots+\lambda_i)^{c_i} \mb{I}_{\lambda_1\ge\ldots\ge \lambda_p\ge 0},
\end{align}
which induces a proper prior if $b_i > 0$, $c_i \ge 0$, for $ i=1,\ldots,p$. Under the model (\ref{BayesModel}), the posterior is
\begin{align} \label{PostLambda}
\pi(\lambda\mid \beta, \sigma^2,y) \propto  e^{- \sum_{i=1}^p (b_j+|\beta|_{(j)}) \lambda_i} \prod_{i=1}^p (\lambda_1+\ldots+\lambda_i)^{c_i+1} \mb{I}_{\lambda_1\ge \ldots \ge \lambda_p\ge 0}.
\end{align}
 The Gibbs sampler can be modified to handle sampling from (\ref{PostLambda}). The conditional posterior distribution of $\lambda_j$ is
\begin{align}
\pi(\lambda_j \mid \lambda_{-j},\beta,\sigma^2,y)					   & \propto e^{- (b_j+|\beta|_{(j)}) \lambda_j} \prod_{i=j}^p (\lambda_1+\ldots+\lambda_i)^{c_i+1} \mb{I}_{\lambda_{j-1} \ge \lambda_j \ge \lambda_{j+1}}.\label{lambdaPostCond}
\end{align}
The conditional posterior (\ref{lambdaPostCond}) can be sampled through rejection sampling using the truncated exponential distribution as the reference distribution. Details are given in Appendix \ref{DetailsGibbs}. 

The hybrid sampler also can be extended to facilitate sampling from (\ref{PostLambda}). Instead of sampling $\lambda$ one coordinate at a time, sample it all at once using Hamiltonian Monte Carlo. The resulting algorithm is described below.
\begin{algorithm}[H]
\caption{The extended block-updating Gibbs Sampler} \label{BlockGibbsExtended}
\begin{algorithmic}[1]
   \item[0:] Fix $T_1, T_2$.
   \item Initialize the parameters $(\beta_{[0]}, \sigma^2_{[0]},\lambda) = (\hat{\beta}_{SLOPE}, \hat{\sigma}^2,\lambda_{BH})$.
   \item Run the Hamiltonian Monte Carlo for time $T_1$, to sample $\beta_{[k]}$ from $\pi(\beta\mid \sigma^2_{[k-1]},\lambda,y)$.
   \item Sample $\sigma^2_{[k]}$ from $\pi(\sigma^2\mid \beta_{[k]},\lambda,y)$.
   \item Run the Hamiltonian Monte Carlo for time $T_2$, to sample $\lambda_{[k]}$ from $\pi(\lambda\mid \beta_{[k]}, \sigma^2_{[k]},y)$.
   \item Repeat 2 through 4 until convergence.
\end{algorithmic}
\end{algorithm}
In algorithm \ref{BlockGibbsExtended}, $\lambda_{BH}$ is the vector of regularization coefficients used by the SLOPE. The Bayesian model with a hyperprior on $\lambda$ is also implemented in STAN modeling language. It can be used along with the STAN package to run algorithm \ref{BlockGibbsExtended} for a generic regression problem, which makes reproducible research more feasible.

\section{Examples} \label{Ex}
\subsection{Simulated data}
\begin{figure}[tbp]
  \centering
  \begin{minipage}[b]{\textwidth}
    \includegraphics[width=\textwidth]{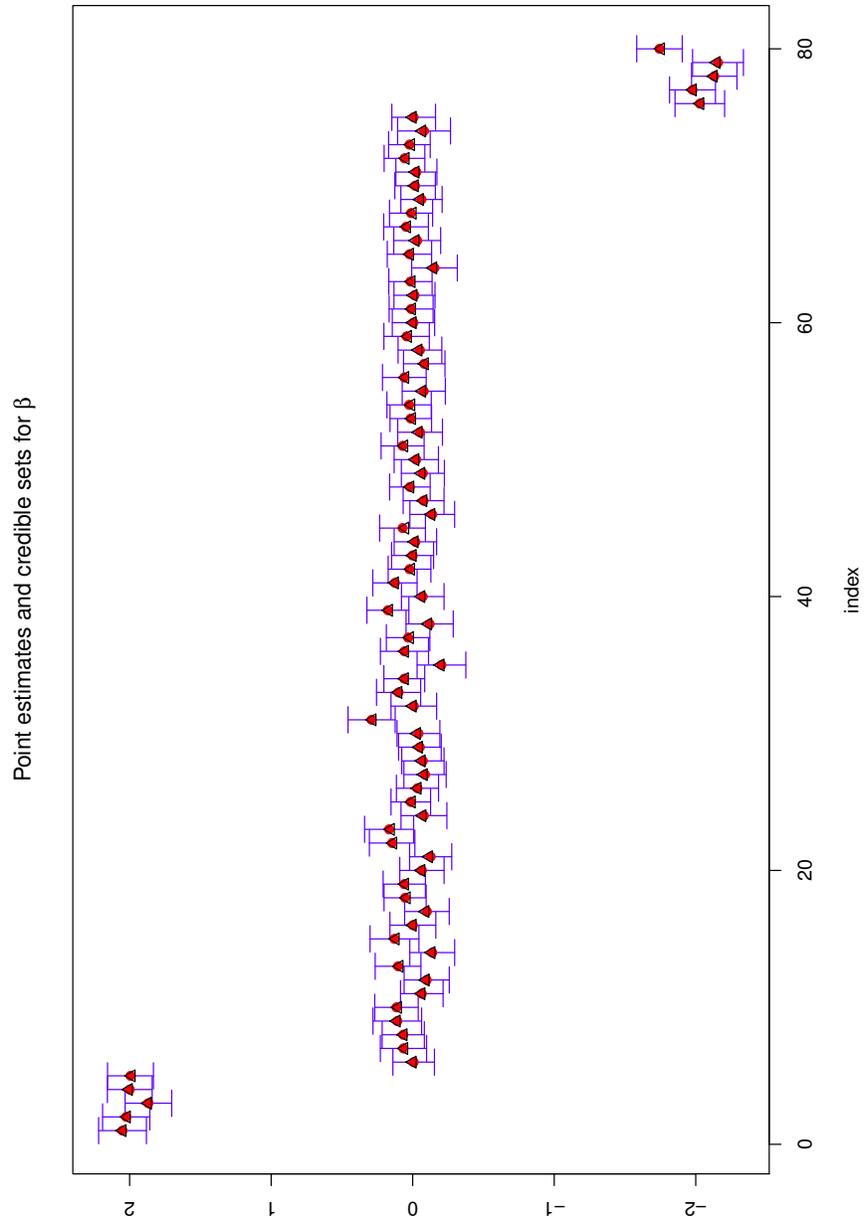}
  \end{minipage}
  \caption{The Bayesian SLOPE posterior mean \textcolor{red}{$\bullet$}, the SLOPE estimate $\triangle$, and $95\%$ Bayesian credible sets for the vector of regression coefficients $\beta$.}
  \label{fig:cred}
\end{figure}

This section compares the SLOPE and the Bayesian SLOPE estimates for simulated data sets. The first experiment involves $200$ observations of $80$ predictors and a response. The design matrix $X$ has independent standard normal entries, the regression coefficients $\beta$ are
\begin{align*}
\beta_i = 2 \text{ for } 1\le i\le 5,\;
 \beta_i = 0 \text{ for } i= 6\le i\le 75,\;
 \beta_i = -2 \text{ for } i= 76\le i\le 80,
\end{align*}
and the errors are standard normal. Both estimates are obtained using the vector of tuning parameters
\begin{align*}
\lambda_i = \Phi(1-\frac{iq}{2p}),  \text{ for } p=80 \text{ and } q = 0.2 .
\end{align*}
The posterior mean is used as the Bayesian point estimate along with the symmetric credible sets. The point estimates along with the $95\%$ Bayesian credible sets are illustrated in Figure \ref{fig:cred}. As can be seen in Figure \ref{fig:cred}, the credible sets cover the true value for most of the variables. There are $5$ non-coverages out of $80$ coefficients, which is expected at the $95\%$ credibility level. The Bayesian SLOPE and the SLOPE estimates agree on all of the coefficients to a great extent. 

The closely matching estimates suggests that the two estimates should behave similarly in predicting the response for future observations. In fact, the Bayesian and empirical Bayes SLOPE estimates, and the SLOPE estimate exhibit similar predictive performance in this example. The out of sample prediction is studied by fitting the three models on a randomly chosen train/test split of the data into groups of 160 and 40 observations; repeated 10 times, using the sum of squares predictive loss function. The estimated prediction errors are presented below in Table \ref{table:pred}. In this simulated data set, the two methods perform similarly in terms of estimation and prediction.

\begin{table}[h!]
\caption{Estimated prediction error}
\label{table:pred}
\centering
\resizebox{\columnwidth}{!}{%
\begin{tabular}{lrrrcrr}
\hline
The SLOPE & \multicolumn{1}{c}{\begin{tabular}{@{}c@{}}The Bayesian SLOPE\end{tabular}} & \multicolumn{1}{c}{\begin{tabular}{@{}c@{}}The empirical Bayes SLOPE\end{tabular}}  \\
\hline
1.151 & 1.166 &  1.197 \\
\hline
\end{tabular}
}
\end{table}

\subsection{Diabetes data set}
\begin{figure}[h!]
  \centering
  \begin{minipage}[b]{\textwidth}
    \includegraphics[width=\textwidth]{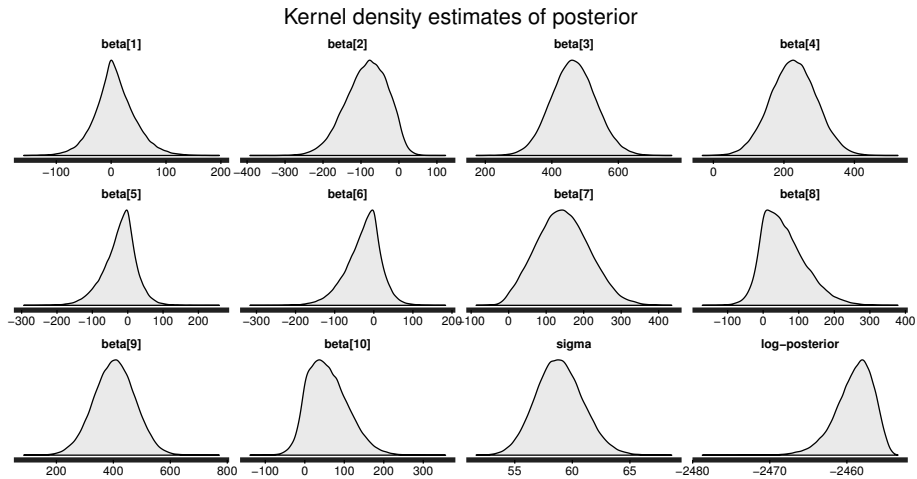}
  \end{minipage}
  \caption{Kernel posterior density estimates for regression parameters. The lower right plot is a kernel density estimate of the log-posterior up to an additive constant.}
  \label{fig:kernel}
\end{figure}

This Section considers the Diabetes data set used by \citet{efron2004least}. The data set includes 442 observations on 10 predictor variables and a response variable. The standardized version of the design matrix has been used. The Bayesian SLOPE has been fitted and compared with the SLOPE and least squares; the result is summarized in Table \ref{table:fit}. Individual kernel posterior density estimates are illustrated in Figure \ref{fig:kernel}.

\begin{table}[h!]
\caption{Estimates of the regression parameters for the diabetes data.}
\label{table:fit}
\centering
\resizebox{\columnwidth}{!}{%
\begin{tabular}{lrrrcrr}
\hline
Parameter & \multicolumn{1}{c}{\begin{tabular}{@{}c@{}}Bayesian SLOPE\\ Mean\end{tabular}} & \multicolumn{1}{c}{\begin{tabular}{@{}c@{}}Bayesian SLOPE\\ Median\end{tabular}} & \multicolumn{1}{c}{\begin{tabular}{@{}c@{}}Bayesian\\ SLOPE SD\end{tabular}}&\multicolumn{1}{c}{\begin{tabular}{@{}c@{}}Bayesian Credible\\ Interval (95\%)\end{tabular}} &\multicolumn{1}{c}{\begin{tabular}{@{}c@{}}SLOPE\\  \end{tabular}} &\multicolumn{1}{c}{\begin{tabular}{@{}c@{}}Least\\ Squares\end{tabular}}  \\
\hline
$\beta_1$ (age) & 6.84 &  4.97 & 36.87 &$ ( -65.87, 85.43)$  & $-6.80$ & $-9.95$\\
$\beta_2$ (sex) &    $-$85.44 &$ -$81.73& 54.66  & $(-200.63 , 7.31)$ & $-235.84$ & $-239.82$\\
$\beta_3$ (bmi) &    465.31 & 464.77 & 66.61 & $(336.36 , 597.38)$ & 522.16 & 519.87 \\
$\beta_4$ (map) &    227.37 &227.26 & 64.88 & $( 100.36 , 354.99)$  & 321.31 & 324.40\\
$\beta_5$ (tc) &    $-$22.51 & $-$17.16& 45.81 &$ (-125.00 , 60.25)$  & $-558.51$ & $-788.31$\\
$\beta_6$ (ldl) &   $-$26.55 & $-$20.57 & 44.68  & $(-127.47 , 51.40)$  & 290.77 & 473.58\\
$\beta_7$ (hdl) &    145.22 & 143.42 & 70.02 & $(15.66 , 286.19) $ & 0.00 & $-99.34$ \\
$\beta_8$ (tch) &     58.77  & 49.41& 61.30 & $(-36.69 , 199.92)$ & 149.21 & 176.70\\
$\beta_9$ (ltg) &    403.61& 404.10 & 72.29 & $(260.21, 543.54)$ & 663.45 & 749.83\\
$\beta_{10}$ (glu) &   58.69 & 53.19 & 50.57 & $(-23.14 , 169.10)$ & 67.41 & 67.60 \\
$\sigma$ & 58.89 & 58.83 &  2.05 & $(55.04 , 63.07)$ &  & \\
 \hline
\end{tabular}
}
\end{table}

\begin{figure}[h!]
  \centering
  \begin{minipage}[b]{\textwidth}
    \includegraphics[width=\textwidth]{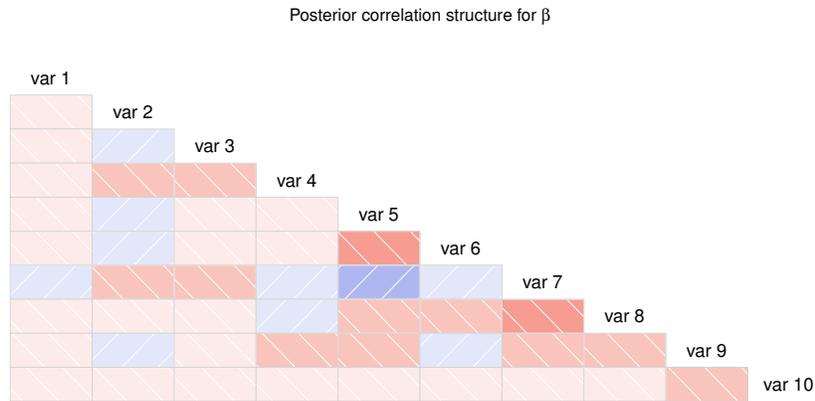}
  \end{minipage}
  \caption{Pairwise posterior correlation between the regression coefficients. Red shows positive correlation and blue shows negative correlation. Darker colors correspond to higher correlation.}
  \label{fig:PairCor}
\end{figure}

The Bayesian SLOPE seems to shrink more than the SLOPE. Interesting, there are some noticeable discrepancies between them for some of the coefficients. 
However, this does not cause conceptual problems because the variables for which there is a significant disagreement are highly correlated. Particularly, we have $corr(X_5,X_6) \approx 0.90$, $corr(X_7, X_8) \approx 0.74$, and $corr(X_6, X_8) \approx 0.66$. It is generally problematic to have highly correlated predictors in the model. Each method estimates differently on the correlated variables. For example, the least squares and the SLOPE provide relatively large values for $X_5$ and $X_6$, with different signs, which cancel out because of the correlation. On the other hand, the Bayesian SLOPE estimates both coefficients with relatively small negative values. A similar effect is present for $X_7$ and $X_8$. The two methods would provide more similar estimates if the correlated pairs were replaced by a linear mixture each. One would expect that highly correlated predictors should result in a posterior with high correlation between corresponding coefficients. This is indeed the case for the Diabetes data set; and can be seen in Figure \ref{fig:PairCor}, which illustrates the pairwise posterior correlations between the regression coefficients.

\begin{figure}[H]
  \centering
  \begin{minipage}[b]{\textwidth}
    \includegraphics[width=\textwidth]{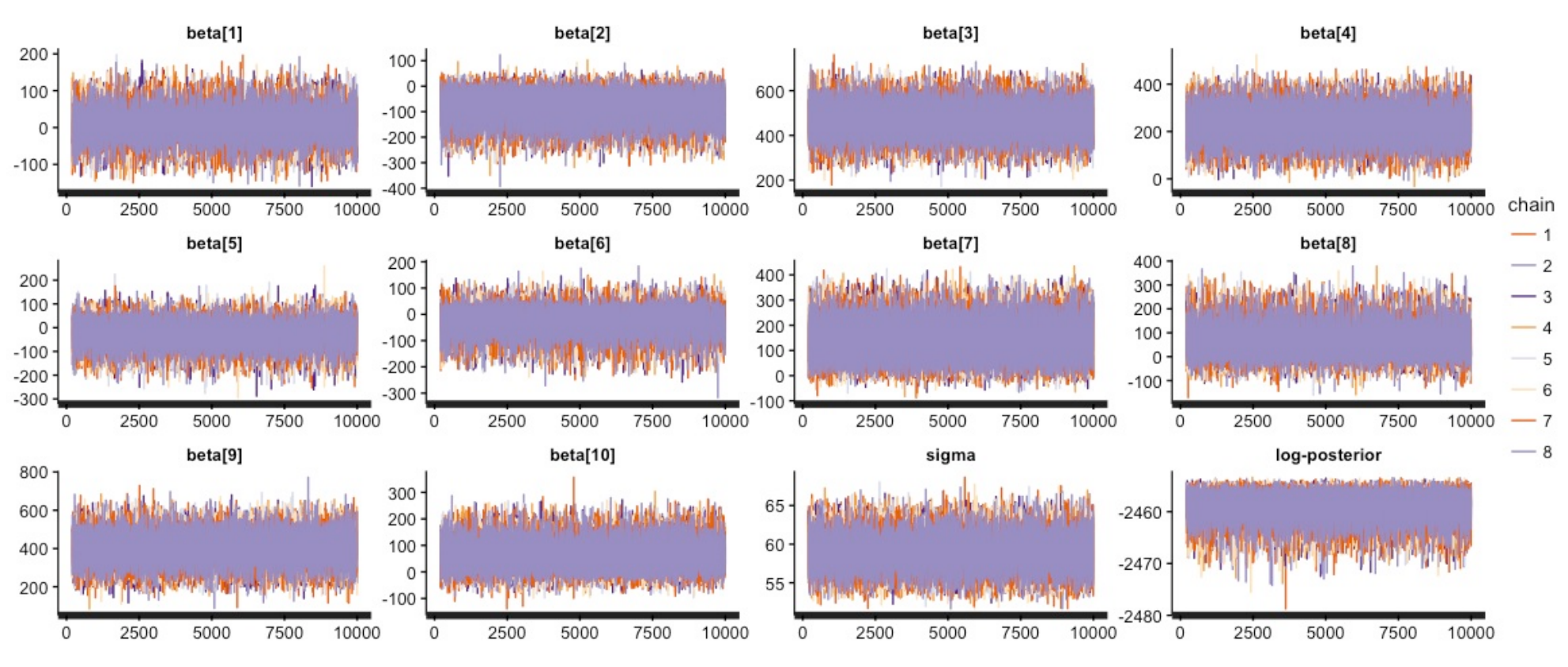}
  \end{minipage}
  \caption{Trace plots for the MCMC sampler, corresponding to different parameters and chains.}
  \label{fig:MCtrace}
\end{figure}

The Hamiltonian Monte Carlo sampler, implemented using STAN, exhibits desirable convergence even after 1000 steps. The results in this Section are obtained based on 10000 steps of 8 parallel chains. For 10000 steps, the lag-three auto-correlation for all the chains is less than $0.02$. A variety of convergence diagnostics are provided in the output from STAN. For instance, Figure \ref{fig:MCtrace} shows the trace plots of the MCMC sampler for the parameters $\beta$ and $\sigma$.

\section{Discussion}
In summary, the Bayesian SLOPE and the SLOPE seem to provide similar estimates with similar predictive performance. The main advantage of the Bayesian SLOPE is access to natural Bayesian credible sets and standard error estimates, whereas there is no natural alternatives for the SLOPE. On the other hand, the SLOPE is faster than the Bayesian SLOPE. The choice between the two depends on the scale of the problem, the computational resources, and the priority of having access to standard error estimates or credible sets.

There are various aspects of the Bayesian SLOPE that could be subject of future investigation.
A possible further direction is to study concentration properties of the posterior (in the sense of \cite{castillo2012needles,van2014horseshoe}).
Another interesting question is the optimality properties of the natural Bayesian estimates, such as the posterior mean or the posterior median. For example, proving minimax optimality for any of these estimators would be of great interest. Applying the Bayesian SLOPE to other real world applications, particularly, to problems in genetics, would be interesting.

\section*{Acknowledgement}
The author is grateful to Cyrus DiCiccio for his helpful comments on the first draft of this paper. The author is supported by a Weiland Graduate Fellowship.

\section*{Supplementary material}
\label{SM}
Supplementary material available online at \url{https://bitbucket.org/amirsepehri/the-bayesian-slope/src} includes R functions and examples, as well as a brief documentation of them.

\appendix

\section{}\label{AppA}
\subsection{Normalizing constant of the SLOPE prior}\label{NormalizingCons}
The normalizing constant, $C(\lambda,\sigma^2)$, for the SLOPE prior $\pi(\beta\mid \sigma^2, \lambda)$ is given by
\begin{align*}
C(\lambda,\sigma^2)^{-1} &= \int e^{\frac{-1}{\sigma} \sum_{i=1}^p \lambda_i |\beta|_{(i)}} d \beta\\
											&= 2^p p! \int_{\beta_1 \ge \beta_2\ge \ldots \ge \beta_p \ge 0} e^{\frac{-1}{\sigma} \sum_{i=1}^p \lambda_i |\beta|_{(i)}} d\beta_1 d\beta_2 \ldots d \beta_p \\
											&=  2^p p! \int_0^\infty e^{\frac{-\lambda_p}{\sigma} \beta_p}  \int_{\beta_p}^\infty e^{\frac{-\lambda_{p-1}}{\sigma} \beta_{p-1}} \ldots \int_{\beta_2}^\infty e^{\frac{-\lambda_{1}}{\sigma} \beta_{1}} d\beta_1 d\beta_2 \ldots d \beta_p .
\end{align*}
Repeated use of $\int_{x}^\infty e^{-c t} d t = \frac{e^{-cx}}{c}$ yields
\begin{align*}
C(\lambda,\sigma^2) &= \frac{\lambda_1(\lambda_1+\lambda_2)\ldots (\lambda_1+\lambda_2+\ldots+\lambda_p)}{2^p \sigma^p p! }.
\end{align*}
\subsection{Unimodality of the posterior}\label{Unimodality}
The argument for unimodality of the SLOPE posterior follows closely from that for lasso \citep{park2008bayesian}. Under the prior  
\begin{align*}
\pi(\beta,\sigma^2) = \pi(\sigma^2) C(\lambda,\sigma^2) e^{\frac{-1}{\sigma} \sum_{i=1}^p \lambda_i |\beta|_{(i)}},
\end{align*}
the joint posterior distribution of $\beta$ and $\sigma^2$ is unimodal in the sense that for all $x$ the upper level set $\{(\beta,\sigma^2)\mid \pi(\beta,\sigma^2)>x, \sigma^2 >0\}$ is connected. To show this, it suffices to show that the posterior is log-concave. This does not hold in the current parametrization. However, the posterior becomes log-concave after a continuous reparametrization (a coordinate transform, not a change of measure). The log-posterior is 
\begin{align*}
\log (\pi(\sigma^2)) - \frac{n+p}{2} \log (\sigma^2) - \frac{1}{2\sigma^2} \|y- X\beta\|^2 - \frac{1}{\sqrt{\sigma^2}} \sum_{i=1}^p \lambda_i |\beta|_{(i)},
\end{align*}
up to an additive term not involving $\beta$ or $\sigma^2$. Define
\begin{align*}
\eta = \beta / \sigma, \;\;\;\;\; \psi = 1/\sigma.
\end{align*}
This is a continuous map with a continuous inverse assuming $0<\sigma^2<\infty$. In $(\eta,\psi)$ coordinates, the log-posterior can be written as
\begin{align*}
\log (\pi(1/\psi^2)) + \frac{n+p}{2} \log (\psi^2) - \frac{1}{2} \|\psi y- X\eta\|^2 -  \sum_{i=1}^p \lambda_i |\eta|_{(i)}.
\end{align*}
The second term is clearly concave. The fourth term is a negated norm, hence concave. The third term is a concave quadratic in $(\eta,\psi)$. Thus, the expression would be concave assuming $\log (\pi(1/\psi^2))$ is concave. Particularly, this holds for the inverse gamma prior and for the scale-invariant improper prior $1/\sigma^2$  on $\sigma^2$. This proves unimodality but not uniqueness of the maximizer. To ensure that maximum is attained uniquely, it suffices to assume that $X$ is full rank and $y$ is not in the column space of $X$ since this makes the quadratic term strictly concave.

\section{}\label{AppB}
\subsection{Details of the Gibbs sampler}\label{DetailsGibbs}
To sample from the marginal posterior of $\lambda$, notice
\begin{align*}
\pi(\lambda_j \mid \lambda_{-j},\beta,\sigma^2,y)					   & \propto e^{- (b_j+|\beta|_{(j)}) \lambda_j} \prod_{i=j}^p (\lambda_1+\ldots+\lambda_i)^{c_i+1} \mb{I}_{\lambda_{j-1} \ge \lambda_j \ge \lambda_{j+1}},\\
																									& \le  e^{- (b_j+|\beta|_{(j)}) \lambda_j} \prod_{i=j}^p (\lambda_1+\ldots+\lambda_{j-2}+2 \lambda_{j-1}+ \lambda_{j+1}+\ldots+ \lambda_i)^{c_i+1}, \\
																									&=   e^{- (b_j+|\beta|_{(j)}) \lambda_j} K(\lambda_{-j}),
\end{align*}
for $\lambda_{j} \in [\lambda_{j+1} , \lambda_{j-1}]$, which can be proved by substituting $\lambda_j$ by $\lambda_{j-1}$ in the product. The last expression can be used for rejection sampling the posterior (\ref{lambdaPostCond}). It suffices to have a method of generating sample from the truncated exponential distribution, which can be done by inverting the cumulative distribution function
\begin{align*}
F(x) = \begin{cases} 0   & x < x_0,\\
 \frac{e^{-c x_0}- e^{-c x}}{e^{-c x_0}-e^{-c x_1}} \;\; &x \in [x_0,x_1],\\
  1 &x > x_1.
\end{cases}
\end{align*}

\bibliography{paper-ref}

\end{document}